\documentclass[fleqn,usenatbib]{mnras}

\usepackage{newtxtext,newtxmath}

\usepackage[T1]{fontenc}

\DeclareRobustCommand{\VAN}[3]{#2}
\let\VANthebibliography\thebibliography
\def\thebibliography{\DeclareRobustCommand{\VAN}[3]{##3}\VANthebibliography}

\usepackage{graphicx}	
\usepackage{amsmath}	
\usepackage{url}
\usepackage{subcaption}
\usepackage{placeins}

\title[Towards detecting non-Gaussianity using CNNs]{Towards detecting Primordial non-Gaussianity in the CMB using Spherical Convolutional Neural Networks}

\author[Jorik Melsen et al.]{Jorik Melsen,$^{1}$\thanks{E-mail: jorik.melsen@gmail.com}
Thomas Flöss,$^{2,3,4,5}$ and
P. Daniel Meerburg$^{4}$
\\
$^{1}$Faculty of Science and Engineering, University of Groningen, Nijenborgh 4, 9747 AG Groningen, The Netherlands\\
$^{2}$Department of Astrophysics, University of Vienna, Türkenschanzstraße 17, 1180 Vienna, Austria\\
$^{3}$Department of Mathematics, University of Vienna, Oskar-Morgenstern-Platz 1, 1090 Vienna, Austria\\
$^{4}$Van Swinderen Institute, University of Groningen, Nijenborgh 5, 9747 AG Groningen, The Netherlands\\
$^{5}$Kapteyn Astronomical Institute, University of Groningen, P.O.Box 800, 9700 AV Groningen, The Netherlands\\
}

\date{Accepted XXX. Received YYY; in original form ZZZ}

\pubyear{\the\year{}}

\begin{document}
\label{firstpage}
\pagerange{\pageref{firstpage}--\pageref{lastpage}}
\maketitle

\begin{abstract}
This paper explores a novel application of spherical convolutional neural networks (CNNs) to detect primordial non-Gaussianity in the cosmic microwave background (CMB), a key probe of inflationary dynamics. While effective, traditional estimators encounter computational challenges, especially when considering summary statistics beyond the bispectrum. We propose spherical CNNs as an alternative, directly analysing full-sky CMB maps to overcome limitations in previous machine learning (ML) approaches that relied on data summaries. By training on simulated CMB maps with varying amplitudes of non-Gaussianity, our spherical CNN models show promising alignment with optimal error bounds of traditional methods, albeit at lower-resolution maps. While we explore several different architectures, results from DeepSphere CNNs most closely match the Fisher forecast for Gaussian test sets under noisy and masked conditions. Our study suggests that spherical CNNs could complement existing methods of non-Gaussianity detection in future datasets, provided additional training data and parameter tuning are applied. We discuss the potential for CNN-based techniques to scale with larger data volumes, paving the way for applications to future CMB data sets.
\end{abstract}




\section{Introduction}

Inflation is one of the cornerstones of modern cosmology.
During a brief period, very early on in the universe's lifetime, shortly after the Big Bang but before nucleosynthesis and the QCD phase transition, the universe underwent an extremely rapid, exponentially accelerated expansion.
Inflation was originally proposed \citep{guth1981inflationary,starobinskii1979spectrum,linde1982new,albrecht1982cosmology} to solve several well-known cosmological conundrums, such as the horizon problem, the flatness problem and a collection of exotic particle problems.
While these problems are indeed explained away by a brief period of exponential expansion, it was soon realized \citep{bardeen1983spontaneous} that quantum perturbations in the inflation potential can source the temperature and polarization fluctuations observed in the cosmic microwave background (CMB). 
The same density fluctuations also evolve into the large-scale structures (LSS) in our universe. 

Despite its successes, the theory of fundamental physics that drives inflation is still ill-understood.
Currently, there is a wide variety of viable mechanisms for driving inflation.
Because inflation happens so early in the lifetime of the universe, it has so far been impossible to probe inflation directly.
Instead of relying on direct measurements of inflation, it is possible to put constraints on inflation through its effects on our observed Universe today. 
Specifically, the statistical properties of the fluctuations we observe in cosmological tracers, such as the CMB, can be related to the statistics of the primordial Universe. 
At present, statistical measurements of the CMB allow us to put the strongest constraints on inflationary parameters.
All viable models of inflation have to be consistent with the Gaussian component of these primordial fluctuations. 
The power spectrum, which captures the Gaussian part of the distribution of these fluctuations, provides only minimal discriminating power between different models of inflation.  
Fortunately, different inflationary scenarios tend to give rise to distinguishable non-Gaussian signatures.
For example, the simplest models of inflation driven by a single scalar field slowly rolling down a potential, predict negligible amounts of primordial non-Gaussianity, whereas models that involve multiple fields can produce observable levels of non-Gaussianity.
The current most stringent constraints on the CMB are consistent with purely Gaussian fluctuations and to a very high significance rule out the presence of a strong non-Gaussian signature. 

While CMB constraints thus rule out any models that predict strong non-Gaussian fluctuations on CMB scales, current constraints are not sufficient to rule out many remaining models, which would require an improvement of bounds by roughly an order of magnitude \citep{meerburg2019primordial,achucarro2022inflation,vazquez2018inflationary}. 
This motivates the ongoing effort to improve these constraints or detect primordial non-Gaussianity, both in the CMB \citep{komatsu2002measurement,cabella2006integrated,yadav2008evidence,yadav2010primordial,ade2014planck,ade2016planck,collaboration2020planck,oppizzi2018cmb} and the LSS \citep{gleyzes2017biasing,slosar2008constraints,ross2013clustering,leistedt2014constraints,ho2015sloan,castorina2019redshift,mueller2021clustering,cabass2022constraints,d2022limits,cabass2022bconstraints,floss2023primordial,floss2024improving}.
To improve over current bounds, there are effectively two approaches. 
The first and obvious option is to improve over current observations, for example by reaching down to smaller scales (CMB) or more dimensions (LSS), both allowing us to sample more independent modes, which will improve our sensitivity. 
Several ongoing and planned experiments are indeed poised to improve constraints by an order of magnitude, at least for some types of non-Gaussianity \citep{SimonsObservatory:2018koc,abazajian2016cmb}.
The other option is to try to improve the analysis that is currently used to obtain the best bounds on non-Gaussianity.
These current methods are based on extracting spatial correlation functions describing the primordial non-Gaussianity present in the CMB.
Among these, the most common statistic to search for signs of non-Gaussianity is the bispectrum, the Fourier equivalent of the three-point correlation function. 
While in principle such spatial correlation functions (or their Fourier equivalents) are optimal for constraining primordial non-Gaussianity from CMB data, their analysis becomes computationally challenging when searching for higher-order primordial non-Gaussianity.
To be able to analyse such types of non-Gaussianity in future data, we thus need to further optimise these methods, or alternatively, consider new methods.
In this paper, we will explore the use of machine learning (ML), or more specifically, spherical convolutional neural networks (CNNs), to extract primordial non-Gaussianity directly from map-level data of the CMB. 
If provided with enough suitable training data, state-of-the-art CNNs might be able to complement, or even compete with, the more traditional methods, especially when searching for signatures that are not captured by the bispectrum alone. 

As far as we are aware, only a limited number of studies have explored the use of ML methods for this purpose. 
One such study is based on another alternative method to analyse non-Gaussianity, relying on the works \citet{hikage2006primordial,hikage2008limits,ducout2013non,buchert2017model} which make use of Minkowski functionals.
These original works compare the precomputed Minkowski functionals of a large set of generated CMB realisations (with varying levels of non-Gaussianity) to the functionals of a target map.
Maximum likelihood estimation then determines the best match and gives us a prediction for the level of non-Gaussianity.
However, in \citet{novaes2015neural} they modify this approach by applying a feed-forward neural network to the Minkowski functionals of the target map instead, where the network directly provides an estimate of the level of non-Gaussianity.
This approach is inspired by a similar ML method, that instead of using Minkowski functionals, feeds wavelet coefficients of the map into a similar neural network \citep{casaponsa2011constraints}.
While this strain of approaches has achieved some success, it has some inherent limitations.
The Minkowski functionals, wavelet coefficients or any other functionals one would use in this manner, are fixed feature extractors, which by design pin down and limit the information that is extracted from the CMB maps.

A more recent study, the only study known to us that deviates from this type of machine learning approach, does try to utilise the flexibility and computational power of more modern ML approaches by making use of CNNs \citet{nagarajappa2024constraining}.
Nevertheless, their approach suffers from some other crucial problems.
In their work, an attempt was made to constrain the level of primordial non-Gaussianity by making flat projections of a set of spherical CMB simulations, using a patch-wise projection algorithm.
On these flat maps, a standard 2D CNN was trained to extract the strength of the non-Gaussianity.
However, projecting the full sky CMB simulations to flat maps in this manner distorts some spatial information from the full sky maps, and removes the spherical symmetry of the data.
We believe, as a result, this method is unlikely to approach the performance of a traditional optimal estimator.
Perhaps even more importantly, a series of implementation mistakes are made in their studies.
This becomes apparent from their discussion of their training and validation data, but also from the results presented in their paper.
These results strongly exceed constraints projected by a simple Fisher analysis of the primordial bispectrum, which represents the theoretically best possible results.
This suggests that their ML model is not reliable and does not give any physical results.

In this paper, we take a different approach, applying a convolutional ML model directly on full-sky CMB maps, such that the ML model has access to all the information it contains.
We achieve this by using specialized spherical CNNs that are trained directly on the CMB sphere.
Additionally, we make use of reliable models and training procedures, ensuring our results are physically sound and can be directly compared to the optimal estimator.
Our aim here is to present a proof-of-concept study, focusing solely on local-type primordial non-Gaussianity, and we compare our results against the traditional bispectrum estimator.
Additionally, we also test our ML models on maps that have been generated with noise as well as maps to which masking has been applied (using masks in line with those normally used to mask out galactic foregrounds).
Finding good agreement, this study motivates further exploration, including application to more exotic models that contain signatures that are not captured by a (separable) bispectrum alone.

The paper is organized as follows. 
In Sec.~\ref{sec:CNN} we describe the two networks we apply to the training data and explain how they differ.
In Sec.~\ref{sec:maps} we discuss how to generate CMB realizations with primordial non-Gaussianity, that serve as the training data for our ML models. 
We briefly describe the experimental details and how we train the network in Sec.~\ref{sec:training}. 
The results are presented in Sec.~\ref{sec:results}, followed by a discussion in Sec.~\ref{sec:discussion} and brief conclusions in Sec.~\ref{sec:conclusions} . 
Several appendices provide some further exploration of the results presented in the main text. 
We have also included an appendix where we apply CNN's to the flat sky, which can be compared to results obtained on the full sky. 

\section{Convolutional Neural Networks}\label{sec:CNN}
Our approach to estimating the strength of non-Gaussianity $f_{\rm{NL}}$ from simulated CMB maps is very different from the traditional methods.
These methods, such as the Komatsu-Spergel-Wandelt (KSW) estimator \citep{komatsu2005measuring}, analyse the statistics of the CMB by applying templates that extract information about the shapes corresponding to the n-point correlation function of a specific type of non-Gaussianity.
This works well for the bispectrum but becomes computationally cumbersome as we go to higher-order n-point correlations.
Here, instead, we apply machine learning, and specifically, neural networks to determine $f_{\rm{NL}}$ directly from the CMB map.
There is a vast amount of different types of neural networks, all with different types of applications.
However, a detailed description of machine learning, neural networks and even the specifics of CNNs is outside the scope of this paper (we refer the reader to e.g. \citet{goodfellow2016deep} for such an introduction).

Instead, here we will focus on giving a brief introduction to CNNs as well as the spherical CNN variations we use.
The reason we want to explore the application of CNNs to CMB map data here is that CNNs excel at image and map-based applications.
A benefit, compared to traditional approaches to extract $f_{\rm{NL}}$, is that we do not need to compute any of the n-point correlations explicitly.
Instead, the network will learn itself what information from the input maps is important to predict $f_{\rm{NL}}$.
This is also the main reason we suspect CNN-based analysis might be able to outperform traditional methods for signals that are not captured by just a (separable) bispectrum.

While CNNs can tackle a variety of different types of Machine learning problems, here we are dealing with a so-called \textit{regression} problem.
A regression problem consists of a dataset $\mathcal{D}=\{\mathbf{x}^i, y^i\}^N_{i=1}$, in our case $\mathbf{x}^i$ are the CMB maps, $y^i$ are the associated true $f^i_{\mathrm{NL}}$, and $N$ is the total number of maps in our dataset.
The goal of the problem is then to learn to predict the correct $f^{i}_{\mathrm{NL}}$ for each input map $\mathbf{x}^i$. 
The trained CNN thus represents a function $F:\mathbf{x}^i\rightarrow \hat{f}^{i}_{\rm{NL}}$.
The better the model performs, the closer the predicted $\hat{f}^i_{\mathrm{NL}}$ will be to the true $f^{i}_{\mathrm{NL}}$. This allows us to define a performance measure for the model, such as the mean squared error (MSE), which we will be using throughout this work.
The way the CNN improves its performance is by iteratively updating a set of trainable parameters by learning from the data points in the dataset.
These trainable neural network parameters (weights) are distributed over various layers.

CNNs generally make use of a few different types of such layers, as well as some layers without any trainable parameters.
The first part of the network, usually called the convolutional part of the network, consists of two types of base layers.
The most important of these are the \textit{convolution layers}.
CNNs have one or more convolution layers, whose weights are contained in a filter that slides over the input map, with the goal of identifying certain features in the data.
The other type instead consists of fixed filters that slide over the maps, known as \textit{pooling layers}. 
The goal of these layers is to compress the data into a lower-dimensional representation.
While a neural network can be purely convolutional, more commonly the convolutional phase is followed by a a set of \textit{fully connected layers}.
The transition between the convolutional phase and the feed-forward phase is given by a \textit{flattening layer}, which transforms the multidimensional output of the convolutional layers into a one-dimensional array.
Then follow one or more fully connected feed-forward layers.
The final layer of a regression CNN is a single output node that represents the target variable, in our case $f_{\mathrm{NL}}$.
One additional thing to note is that the output of each trainable layer (except for the final output layer) applies a non-linear \textit{activation function}, which increases the predictive capabilities of the network.
In this work, we will be using the leaky ReLU function, a derivative of the well-known ReLU function, and is given by

\begin{equation}
    \text{Leaky ReLU}(\mathbf{z})=
    \begin{cases}
        b\mathbf{z} & \text{if } z<0, \\
        \mathbf{z} & \text{if } z\geq0,
    \end{cases}
\end{equation}

\noindent where we will use $b=0.3$.

To be able to update the weights, $\mathbf{w}$ we need to define a \textit{loss function}.
In plain terms, the loss function guides the training process of the network in the right direction.
More formally, the loss function gives us an expression for how well the output of the CNN $\hat{y}^{i}$ agrees with the true value $y^{i}$, for a given input $\mathbf{x}^{i}$ and the current set of model weights $\mathbf{w}$.
To actually train the network, the weights can be updated by performing \textit{stochastic gradient descent} on the loss function.
We will use the squared error for our loss function, which gives

\begin{equation}
    \mathcal{L}(\mathbf{x}^i, y^i, \mathbf{w}) = (\hat{y}^{i}(\mathbf{x}^i, \mathbf{w})-y^i)^2,
\end{equation}

\noindent In our case, $\mathbf{x}^i$ represents the $i$-th CMB map in our data, $y^i$ represents the associated true $f^i_{\mathrm{NL}}$ and $\hat{y}^i$ is the $\hat{f}^{i}_{\mathrm{NL}}$ predicted by the network, given this map.
The stochastic gradient descent update for any weight parameter in the model can then be defined as

\begin{equation}\label{eq:GDLI}
    w'_i=w_i-\eta\frac{\partial\mathcal{L}(\mathbf{x}, y, \mathbf{w})}{\partial w_i},
\end{equation}

\noindent where $\eta$ is the learning rate, a parameter that controls the step size of each gradient update, and the gradient itself is calculated by back-propagating through the network.
In reality, it is more common to update the weights for several data points at the same time, which is known as \textit{batch gradient descent} or \textit{mini-batch gradient descent}.
In this work, we will be making use of batch updates as well.
Where the weights for each training batch are thus updated by applying the MSE to the input maps in that batch.
Additionally, more advanced methods to update the weights exist as well (collectively generally referred to as \textit{optimization algorithms} or \textit{optimizers}, which also includes standard gradient descent).
For example, instead of simply using the gradient, some optimizers uses weighted momentum terms, which do not just depend on the current gradient, but also the gradient of previous update steps.
One such optimizers is the well know \textit{Adam optimizer}, which makes use of two distinct momentum terms and also uses a dynamic learning rate \citep{kingma2014adam}.
We will be using the Adam optimizer in this work.

\subsection{Spherical CNNs}
Finally, we have to deal with one additional complication compared to standard CNN applications.
Conventional CNNs are designed for applications that use data formatted on a regular grid (e.g. flat 2D images, video data, volumetric data with cubic pixels, etc).
However, in our case, we are dealing with spherical full-sky CMB maps, on which these standard CNNs cannot be straightforwardly applied.
While there are multiple ways to discretize a spherical surface, there is, unfortunately, no known method that results in perfect uniform sampling \citep{mcewen2011novel,driscoll1994computing,beentjes2015quadrature}.
As is common in cosmology, we adopt the Hierarchical Equal Area isoLatitude Pixelation (HEALPix) scheme to create our pixelation of the spherical maps \citep{gorski2005healpix}.
This scheme has three key properties, it is hierarchical (under up- and downsampling), all pixels have an approximately equal area and shape, and the pixels are organised in rings of the same latitude.
The former two of these properties simplify the operations necessary for our CNN, such as the pooling and convolutions.
Nevertheless, it is not possible to straightforwardly apply a standard CNN to these maps due to two main problems.
First, the number of neighbours is not the same for all pixels (most have 8, but some have only 7), which makes it non-trivial to apply a fixed-sized filter without modifications.
And second, there is no clear beginning or end to the sphere, so we have to be careful to make sure we only convolve every neighbourhood of pixels once.

Here we will adopt two methods that have been successfully applied to the HEALPix sphere.
We refer to the first method as the pixel-based HEALPix CNN, developed by \citet{krachmalnicoff2019convolutional}.
It works very similar to a classic CNN.
Following the HEALPix pixel ordering, it performs convolutions on the sphere by applying a filter to each pixel and its direct neighbours once.
The pixels that only have 7 instead of 8 neighbours are padded with an extra neighbour with a value of 0.
Unlike a standard 2D CNN, the pixel-based CNN makes use of 1D convolutions, where each pixel and its neighbours are flattened into a 1D array, and the convolution is taken over this array.
For the pooling layers, the hierarchical nature of the HEALPix scheme allows for recursive downsampling of 4 pixels into their superpixel, until the base HEALPix sphere is reached, which only has a total of 12 pixels.

While this method works reasonably well, it also has a few downsides.
First, padding missing pixels with a zero will create a bias towards zero in the convolution of pixels with missing neighbours.
While this can be circumvented by e.g. padding with the average of the 8 other pixels (the pixel itself and its 7 neighbours), this makes the CNN more computationally inefficient.
Second, and perhaps even more important, is that this CNN does not preserve rotational invariance. 
Because the pixels are convolved in a set order, filters are oriented in a fixed way, if we rotate the sphere, the convolution will be applied to the sphere in a different orientation.
This means the outcome of the convolution will be different and valuable properties of the sphere are lost.

The second method we adopt does not suffer from these issues.
The DeepSphere CNN does not perform convolutions directly on the HEALPix maps \citep{perraudin2019deepsphere}. 
Instead, it creates a graph representation of the maps on which the convolutions are performed.
These convolutions, as opposed to the convolutions of the pixelized HEALPix CNN as well as standard CNNs, are radial in nature.
What this means is that the convolved node itself is multiplied by a weight $w_{0}$, the direct neighbours of the convolved node are all multiplied by the same weight $w_{1}$, the next radial layer of the closest indirect neighbours are all multiplied by a weight $w_{2}$, etc.
The graph thus naturally divides the nodes around a pixel in rings of the same radius, each such ring being multiplied by a single unique weight.
Not only does this reduce the number of trainable parameters (e.g. a $3 \times 3$ convolution of the pixelized CNN with 9 weights is equivalent to only 3 radial weights in the DeepSphere CNN), but it also allows for a more flexible choice of kernel size, as it is not hindered by missing neighbours.
Furthermore, it is equivariant or invariant under rotations (depending on the exact application and hyperparameter settings). 
Pooling is handled the same way in DeepSphere as in the pixelized CNN (a new graph being computed for the reduced nside), although DeepSphere also provides more involved weighted downsampling schemes.

\section{Generating CMB maps}\label{sec:maps}
As mentioned previously, in this work we will be focusing on inflationary models that lead to primordial non-Gaussianity of the local type.
We can define a parametrization of the primordial curvature perturbation $\Phi(\mathbf{x})$ by splitting it up in its Gaussian components $\Phi_{\rm{L}}(\mathbf{x})$ and it's non-Gaussian (and non-linear) component $\Phi_{\rm{NL}}(\mathbf{x})$ as follows \citep{verde2000large,komatsu2001acoustic}:

\begin{equation}\label{eq:non-gaus}
    \Phi(\mathbf{x})=\Phi_{\rm{L}}(\mathbf{x})+f_{\rm{NL}}\Phi_{\rm{NL}}(\mathbf{x}),
\end{equation}

\noindent where $f_{\rm{NL}}$ here indicates the strength or amplitude of the non-Gaussianity, with $f_{\rm{NL}}=0$ corresponding to purely Gaussian perturbations. $r$ is the conformal distance.
For non-Gaussianity of the local type, the non-Gaussian component is quadratically dependent on the Gaussian component  

\begin{equation}\label{eq:local-non-gaus}
    \Phi_{\rm{NL}}(\mathbf{x})=\Phi^2_{\rm{L}}(\mathbf{x})-\langle\Phi^2_{\rm{L}}(\mathbf{x})\rangle.
\end{equation}
We are interested in determining the value of $f_{\rm{NL}}$ from the CMB, which allows us to evaluate the feasibility of certain theories of inflation, e.g. an observable $f_{\rm{NL}}$ value favours a multi-field model, while $f_{\rm{NL}}=0$ is consistent with a single-field model.
To train, our network requires a large number of valid CMB realizations with varying $f_{\rm{NL}}$ values.
There are multiple methods to generate such maps \citep{komatsu2003first,liguori2003high,liguori2007temperature,elsner2009improved,smith2011algorithms}. Although the method in \citet{smith2011algorithms} appears to be the most efficient, we note that it has been designed to generate only the correct two- and three-point statistics in the map. Since we aim to use the full map-level data, we will use maps generated using the approach of \citet{liguori2003high,liguori2007temperature,elsner2009improved}, which by construction are correct at all orders of statistics. We briefly outline this procedure below:
\begin{enumerate}
    \item Generate Gaussian harmonic coefficients $\Phi_{\rm{L};\ell m}(r)$ of the primordial perturbations, along the line-of-sight $r$ to the surface of last-scattering.
    \item Take the spherical harmonic transform to obtain the perturbations in real (pixel) space $\Phi_{\rm{L}}(\hat{n},r)$, where $\hat{n}$ denotes the position (angle) on the sky.
    \item Obtain the non-Gaussian component $\Phi_{\rm{NL}}(\hat{n},r)$ using equation \eqref{eq:local-non-gaus}.
    \item Perform the inverse harmonic transform to obtain the non-Gaussian expression $\Phi_{\rm{NL};\ell m}(r)$ in harmonic space.
    \item  Obtain the harmonic coefficients of the CMB temperature for both the Gaussian and non-Gaussian perturbations
    \begin{equation}
        a^X_{\ell m}=\int dr \; r^2\Phi_{\ell m}(r)\alpha^T_{\ell}(r),
    \end{equation}
    where 
    \begin{equation}
        \alpha^X_{\ell}(r)=\frac{2}{\pi}\int dkk^2g^X_{\ell}(k)j_{\ell}(kr),
    \end{equation}
    is the real-space transfer function, $g^{X}_{\ell}$ is the Fourier-space transfer function and $X$ denotes the type of CMB signal (temperature or polarization).
    \item Similar to equation \eqref{eq:non-gaus}, the CMB signal including local primordial non-Gaussianity can then be obtained using
    \begin{equation}\label{eq:alm}
    a^T_{\ell m}=a^T_{\rm{L};\ell m}+f_{\rm{NL}}a^T_{\rm{NL};\ell m}.
\end{equation}
\end{enumerate}
We can then take the spherical harmonic transform of $a^T_{\ell m}$ to obtain a map of the temperature fluctuations in the CMB.
Given the computationally demanding nature of generating these $a^T_{\ell m}$, we here use a set of 1000 $a^T_{\rm{L};\ell m}$ and $a^T_{\rm{NL};\ell m}$ pairs generated by \citet{elsner2009improved} to obtain our maps.

\begin{table*}
    \centering
    \begin{tabular}{|c||c|c|c|c|c|c|}
        \hline
         & \multicolumn{2}{c|}{DeepSphere} & \multicolumn{2}{c|}{Pixel-based} & \multicolumn{2}{c|}{KSW}\\
        \hline
        \hline
        nside & Full range & Gaussian & Full range & Gaussian & Full range & Gaussian\\
        \hline
        \hline
        16 & 183 & 206 & 186 & 214 & 206 & 189\\
        \hline
        32 & 95 & 103 & 101 & 106 & 126 & 93\\
        \hline
        64 & 56 & 52 & 58 & 54 & 104 & 47\\
        \hline
        128 & 33 & 28 & 35 & 31 & 81 & 22\\
        \hline
    \end{tabular}
    \caption{Results for the best performing models of both the DeepSphere architecture and the Pixel-based architecture that were trained on the unmodified dataset. The RMSE is given for the full range test set, containing $f_{\rm{NL}}$ uniformly distributed over the range $-1000$ to $1000$, and for the Gaussian test set (where all maps have $f_{\rm{NL}} = 0$). The KSW estimator RMSE is also given for both of these test sets. Note here that the KSW estimator is only optimal at $f_{\rm{NL}}=0$, away from $f_{\rm{NL}}=0$ the estimator becomes increasingly suboptimal, as can be seen by the large RMSE on the full range test set. }
    \label{tab:normal}
\end{table*}

\begin{table}
    \centering
    \begin{tabular}{|c||c|c|c|c|c|}
        \hline
        Model & \multicolumn{2}{c|}{DeepSphere} & \multicolumn{2}{c|}{Pixel-based} & Bispectrum\\
        \hline
        \hline
        nside & Full range & Gaussian & Full range & Gaussian & Gaussian\\
        \hline
        \hline
        16 & 207 & 236 & 219 & 242 & 219\\
        \hline
        32 & 105 & 113 & 110 & 119 & 105\\
        \hline
        64 & 61 & 53 & 61 & 61 & 53\\
        \hline
        128 & 34 & 29 & 38 & 35 & 25\\
        \hline
    \end{tabular}
    \caption{Results for the best performing models of both the DeepSphere architecture and the Pixel-based architecture that were trained on the noisy dataset. The RMSE is given for the full range test set, containing $f_{\rm{NL}}$ uniformly distributed over the range $-1000$ to $1000$, and for the Gaussian test set (where all maps have $f_{\rm{NL}} = 0$). The bispectrum Fisher error is also given (at $f_{\rm{NL}}=0$).}
    \label{tab:noise}
\end{table}

\begin{table}
    \centering
    \begin{tabular}{|c||c|c|c|c|c|}
        \hline
        Model & \multicolumn{2}{c|}{DeepSphere} & \multicolumn{2}{c|}{Pixel-based} & Bispectrum\\
        \hline
        \hline
        nside & Full range & Gaussian & Full range & Gaussian & Gaussian\\
        \hline
        \hline
        16 & 223 & 247 & 225 & 252 & 236\\
        \hline
        32 & 118 & 128 & 121 & 129 & 116\\
        \hline
        64 & 67 & 61 & 68 & 67 & 59\\
        \hline
        128 & 37 & 33 & 41 & 36 & 28\\
        \hline
    \end{tabular}
    \caption{The same applies as in the caption of table \ref{tab:noise}, the only difference being that now training and testing was done on the masked dataset.}
    \label{tab:mask}
\end{table}

\section{Setting up the training data and the network}\label{sec:training}
Because our CNNs will require a large amount of CMB maps to train on, we have first generated a dataset of 12000 maps for nside 16, 32, 64 and 128, so that we can train and test the model on different resolutions.
10000 of these are training maps, 1000 are validation maps and 1000 are test maps.
For the training maps, 400 of the $a^T_{\ell m}$ seeds from \citet{elsner2009improved} were used.
The Gaussian and non-Gaussian components of each seed were used to create 25 distinct $a^T_{\ell m}$ realisations, each receiving a random rotation and a random $f_{\rm{NL}}$ in the range -1000 to 1000 (using equation \eqref{eq:alm}).
The validation set was generated similarly from a different 100 seeds (each seed being used for 10 maps, all with their own random rotation and $f_{\rm{NL}}$).
For the test set, however, we used the remaining 500 seeds for only 1000 maps to make the test set more independent, and therefore representative for testing.
Where each seed was used to generate 2 maps with a random $f_{\rm{NL}}$ (again drawn from the range -1000 to 1000), but no rotations were applied.
It is important here to stress that we used different seeds for the training, validation and test sets, so that they are all independent of each other.
Additionally, we have also created a purely Gaussian test set, consisting of 10000 entirely unique maps, using the same power spectrum that was used to generate the $a^T_{\ell m}$ from \citet{elsner2009improved}.

We also modified the aforementioned complete dataset, creating two additional datasets in the process.
The first of these was created by adding noise to all maps. 
A separate realisation of the noise was generated for each map, using the following noise power spectrum
\begin{equation}
    C_{\rm{noise}}(l)=N^2_{\rm{inst}}\times\left(1+\left(\frac{l_{\rm{knee}}}{l}\right)^{\alpha_{\rm{knee}}}\right),
\end{equation}
where $N_{\rm{inst}}$ represents the power spectrum of instrumental noise, and $l_{\rm{knee}}$ and $\alpha_{\rm{knee}}$ are parameters that control the position and the tilt of the power spectrum \citep{barron2018optimization}.
In this analysis, we have used the following parameters to generate the noise: $N^2_{\rm{inst}}=8\times10^4\;(\mu\text{K arcmin})^2$, $l_{\rm{knee}}=100$ and $\alpha_{\rm{knee}}=3$.
Note that these parameter settings are not based on any real data and are notably larger than the noise levels in current and future observational data (e.g. Planck, SO and CMB-S4).
This high level of noise was chosen to ensure the noise has a measurable effect on the performance of the optimal estimator at the relatively low resolutions we are investigating.
The second modified dataset was created by applying a mask to the original data.
We used the type of mask that is typically used to block out the galactic plane from the CMB.
Specifically, we use the mask from \citet{masks} where the visible fraction of the sky is equal to 0.8.

To train on this data, we use the spherical networks described in the previous section.
For both networks, we use a similar architecture, the only difference between the two being the way they perform spherical convolutions;
the pixel-based network using 1D convolutions in real space, that operate similarly to the 2D convolutions of a standard flat CNN, while DeepSphere uses radial graph-based convolutions (where we have used the setting $k=2$).
The architecture consists of a convolutional phase, followed by a feedforward phase.
The convolutional phase contains a number of convolution blocks, each block starting with a convolution, followed by a dropout layer with a rate of $0.1$ and then an average pooling layer.
The number of convolutional blocks is determined by the nside of interest: for each nside we use the number of blocks that are necessary to pool down to nside 1, e.g.
nside 16 requires 4 blocks ($16\rightarrow8\rightarrow4\rightarrow2\rightarrow1$) and for every step up in nside, we added another block.
After the convolutional phase, we apply a flattening layer followed by a dropout layer with a rate of 0.3.
The feed-forward phase consists of 2 feed-forward layers with 32 nodes each and a final output layer, giving us the predicted $f_{\rm{NL}}$ value.
Throughout the entire network, we use Leaky ReLU for activation, noting that the output layer does not use any activation.
On top of this architecture, we use a fixed set of other hyperparameters.
The loss function is, as mentioned before, the MSE loss between predicted and true $f_{\rm{NL}}$, and we use the Adam optimizer to train our network. This optimizer also includes weight decay, with a decay rate set to 0.1.

\section{Results}\label{sec:results}
\begin{figure*}
    \centering
    \begin{subfigure}[t]{0.48\linewidth}
        \centering
        \includegraphics[width=\linewidth]{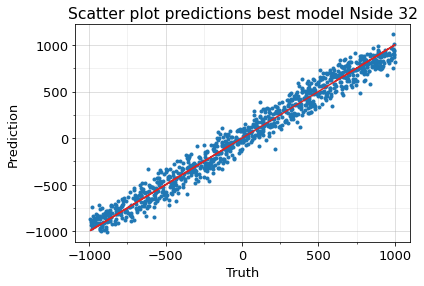}
        \caption{Plot for the model trained on nside 32 maps.}
        \label{fig:scatter_nside32}
    \end{subfigure}\hfill%
    \begin{subfigure}[t]{0.48\linewidth}
        \centering
        \includegraphics[width=\linewidth]{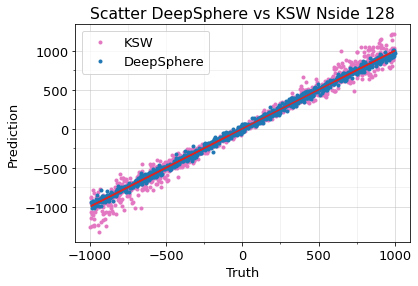}
        \caption{Plot for the model trained on nside 128 maps. Additionally here we have also displayed the scatter for the $f_{\rm{NL}}$ predicted by KSW estimator on the test set.}
        \label{fig:scatter_nside128}
    \end{subfigure}
    \caption{Scatter plots of the predicted $f_{\rm{NL}}$ values of the best DeepSphere model on the full range test set (which contains $f_{\rm{NL}}$ values uniformly distributed from $-1000$ to $1000$) versus the correct $f_{\rm{NL}}$ values. The red lines here represent a hypothetical flawless estimator.}
\end{figure*}

For all tested settings, both network types were able to learn to predict $f_{\rm{NL}}$ from generated CMB maps.
Here we will be focusing on the results from the best test models, with the full set of results presented in the Appendix \ref{app:full-results}.
While the CNNs were not able to perform optimally for all settings, for the best DeepSphere models on nside 16 till 64, all Gaussian test errors were within 10\% of the optimal error, and for nside 128, all errors are within 24\%.
Comparably but slightly worse, for the best pixel-based models on nside 16 to 64, all Gaussian test errors were within 14\% of the optimal error, and for nside 128, all errors are within 34\%.
In Table \ref{tab:normal} we show the detailed results for the best-performing models on the dataset without noise or masking.
Here we display the results for the DeepSphere CNN and the pixel-based CNN. 
We also show the error obtained from an optimal estimator, where we specifically use the KSW estimator implemented by \citet{AdriKSW}.
For both the CNN models and the optimal estimator, we display both the RMSE on the test set that contains maps with an $f_{\rm{NL}}$ drawn uniformly between $-1000$ and $1000$, as well as the more extensive test set with $f_{\rm{NL}}=0$.
For all tested nsides, DeepSphere performs better than the pixel-based network.
Nevertheless, both networks show a similar trend when compared to the optimal error.
For nside 16 both networks perform around the the optimal Gaussian error over the entire range of tested $f_{\rm{NL}}$ values for the full range test set, but for higher nside it starts to increasingly diverge away from the optimal error at $f_{\rm{NL}}=0$.
Results for the Gaussian test set show an opposite trend. 
For low nside the error at $f_{\rm{NL}}=0$ is noticeably higher than the optimal error, while for increasing nside the RMSE seems to get closer to the optimal error.
From nside 16 to 64, the difference between the experimental error and the optimal error seems to shrink equivalently to the magnitude of the errors.
However, for nside 128 this difference seems to go up again.

Aside from the RMSE, another important metric that we haven't discussed so far is the magnitude of any bias in the predictions of the CNN models.
Whereas by construction the optimal KSW estimator is unbiased at $f_{\rm{NL}}=0$, the CNN models do show a small non-zero bias at $f_{\rm{NL}}=0$.
For example, for the normal test maps, the best DeepSphere model respectively has a bias of -5, 5, 2 and 1 for nside 16, 32, 64 and 128, respectively.
This is relatively small compared to the magnitude of the Gaussian RMSE, and the bias seems to shrink proportionally with an increase in nside.
For the full bias results, we refer the reader again to Appendix \ref{app:full-results}.

It is also worth looking at an example of a scatter plot produced by one of the CNN models, to get a better understanding of how the CNN predictions behave over the full range test set.
In figure \ref{fig:scatter_nside32} we have displayed the scatter plot of the best performing DeepSphere model for nside 32.
We see that across almost the entire range of $f_{\rm{NL}}$ values in the test set, the model has a consistent error, displays very little deviations in the variance of the predictions and it shows no clear consistent bias.
However, close to the $|f_{\rm{NL}}|=1000$, predictions show a noticeable tail effect, having a bias towards lower $f_{\rm{NL}}$ and a collapse in the variance (which results in the RMSE being similar at this range compared to the smaller $f_{\rm{NL}}$ values). 
This tailing effect does get suppressed as the nside increases, as can also be seen in figure \ref{fig:scatter_nside128}, which shows the scatter for the best DeepSphere model on nside 128.

Additionally, although the optimal estimator is by design only optimal around $f_{\rm{NL}}=0$, it is worth highlighting the difference between the CNNs and the bispectrum estimator over the full test range.
As can be seen in table \ref{tab:normal}, both CNNs perform a lot better than the optimal estimator when evaluated on the full test range, this difference only becoming large with an increasing resolution.
In figure \ref{fig:scatter_nside128} we have also plotted the predictions of the KSW estimator on the test set, allowing for a more in depth comparison.
We see that for small $f_{\rm{NL}}$ values, as expected, the optimal estimator performs better than DeepSphere.
As the $f_{\rm{NL}}$ values increase, however, the performance of KSW becomes increasingly worse, whereas the DeepSphere model's performance remains consistent.

The results of the best-performing models for maps with noise are given in Table \ref{tab:noise} and for masked maps in Table \ref{tab:mask}, where we have also displayed the Fisher forecast of the optimal error achievable with a bispectrum/KSW analysis for both cases.
These results seem to show similar behaviour for a change in nside as the results in table \ref{tab:normal}.
The Gaussian error seems to get increasingly closer to the optimal error, but then again increases for nside 128.
Overall, both spherical CNNs seem to work for both masked and noisy data.
In particular, the network seems to perform relatively better on the noisy maps than on the noiseless maps.
Note that for noisy maps with nside 64, the Gaussian RMSE for DeepSphere even performs on par with the optimal estimator.

\section{Discussion} \label{sec:discussion}
The primary purpose of this work was to provide a proof of concept for the use of spherical CNNs to extract $f_{\rm{NL}}$ from CMB temperature maps.
Our results are promising and show that spherical CNNs are indeed able to learn to determine $f_{\rm{NL}}$ given a training set of simulated maps.
Both the DeepSphere and pixel-based CNNs were able to give a relatively accurate prediction of $f_{\rm{NL}}$ on an independent test set after training was completed.
However, the DeepSphere network performed better than the pixel-based network in every tested scenario.
Aside from its success, we do see that with the current hyperparameter settings, DeepSphere does not perform optimally in most scenarios.
While for the full range test set, the network performed close to the optimal error for nside 16 and 32, the Gaussian test is only near on par with optimal for the noisy maps with an nside of 64.
Furthermore, as can be seen in appendix \ref{app:full-results}, there is a non-negligible variation in performance between different model training runs, the above only focusing on the best of all runs.

One important point to make is that the optimal KSW estimator is only optimal at $f_{\rm{NL}}=0$, the error going up strongly as we get to large $f_{\rm{NL}}$.
On the other hand, the CNN shows that it performs relatively consistently over the full range of $f_{\rm{NL}}$.
For lower nside, however, this might work to its detriment.
Given that the errors are large for low nside, granting the network some freedom in what parameters result in the best overall performance, it seems that in an effort to keep the error low over the full range of $f_{\rm{NL}}$, this ends up hurting the performance for low $f_{\rm{NL}}$.
When the error bounds get tighter, for larger nside, this problem seems to be alleviated, the Gaussian error approaching the optimal error.
Additionally, we also see some edge effects in the predictions of the CNNs near $|f_{\rm{NL}}|=1000$.
However, this is merely due to the network not seeing examples of maps with $|f_{\rm{NL}}|>1000$ during training.
Increasing or decreasing the range of $f_{\rm{NL}}$ in the dataset shifts this tail effect with it.
If we then ensure the range of $f_{\rm{NL}}$ is sufficiently large for the application, these edge effects should not cause any issues.
Furthermore, with an increase of the nside, the tailing effect also becomes increasingly smaller.
This is likely also because of the deceasing overall error, such that the error around $|f_{\rm{NL}}|=1000$ becomes sufficiently small for the models to naturally be able to predict $|f_{\rm{NL}}|>1000$ within the error margin (because of the stochastic nature of the training process), and an absence of maps with $|f_{\rm{NL}}|>1000$ does therefore no longer hinder its performance.
Finally, the full range test set only contains a limited number of independent examples, and we expect that with a greater test set, we would get a more accurate error.

For the Gaussian test error, which is the test we should compare to the optimal estimator, it seems that from nside 16 to nside 64 there is a positive trend, the test error shrinking proportionally or even reaching optimal performance (for the noisy maps).
This trend is however broken for nside 128.
The reason for this is likely related to the reason why the CNNs seem to perform better on noisy maps.
For higher nside, we add another layer, which adds additional parameters to the network.
The larger the network, the more independent training data is required to fine-tune all parameters.
Given that the data only contains 10000 test maps, generated from only 400 independent seeds, it seems the training set is not independent/large enough to realise the full potential of the CNNs for larger nside (and possibly also for lower nside).
This naturally explains the better performance for noisy data.
Adding noise to the network, similar to rotations, adds some extra variation to the data, which allows the network to better train on the relevant underlying statistics.

To test this hypothesis, we have performed an additional test for all nside maps without noise and masks.
Instead of generating only 10000 training maps, we have generated 40500 training maps for this second experiment, using 900 of the independent $a_{\ell m}$ from \citet{elsner2009improved}, reserving the final 100 for the validation set and using the same Gaussian test set as before (but no test over the full range of $f_{\rm{NL}}$, see appendix \ref{app:large-training-results} for the full details and the results).
This analysis indeed seems to indicate that more data is required to fine-tune the parameters of the CNNs for higher nside.
The performance on nside 16 is fairly similar, but the performance is better for all other nside, even reaching optimal performance for nside 64.
It is therefore expected, that with even more (independent) training data, the CNN will be able to reach optimal performance on nside 128 as well.

The question then arises as to why the network is able to reach optimal performance for higher nside, but not for lower nside. This is likely due to the very same underlying mechanism; the CNNs for lower nside have fewer layers and parameters and are therefore less computationally powerful.
This is, however, not to say that a different type of spherical CNN than the ones we have discussed here would not be able to perform optimally on lower nside and/or higher nside with less training data.
Spherical CNNs are a relatively new field of research, with new architectures being reported on regularly.

To eliminate the current limitations of spherical CNNs, we also performed a simplified test with flat CNNs, which benefit from a much more extensive history of research.
While applying CNNs to projections of the full sky simulations distorts a lot of valuable information (and leads to performance worse than the spherical CNNs), we can generate non-Gaussian flat maps that are inherently flat by avoiding the need to use spherical harmonics altogether, using the Fourier domain instead.
The full description of this analysis and the results can be found in appendix \ref{app:flat-results}.
In this simplified flat scenario, we can reach optimal performance even on the relatively low test resolution.
This supports the argument that a different class of spherical CNNs might be able to achieve optimal performance for any nside. 

Finally, even when the error of CNNs is seemingly on par with the optimal estimator, this is only meaningful if, like for the optimal estimator, the predictions are unbiased.
We observed that the CNNs in most cases only have a very small bias, or in some cases, no bias, relative to the magnitude of the error.
Any bias that is present can also straightforwardly be subtracted from the predictions of the models, as discussed in more detail in Appendix \ref{app:full-results}, resulting in unbiased CNN predictions.
This also very slightly improves the performance of the models, but overall only has a marginal effect, highlighting that the errors we observe are directly comparable to the optimal estimator.
This further supports the notion that spherical CNNs can perform on par with the optimal estimator.

\section{Conclusions}\label{sec:conclusions}
In this paper, we demonstrate that spherical convolutional neural networks (CNNs) offer a promising new approach for detecting primordial non-Gaussianity in the cosmic microwave background (CMB). 
Our results, obtained from both DeepSphere and pixel-based CNN models, indicate that CNNs can approximate traditional optimal error bounds when trained on full-sky CMB maps under various conditions, including noisy and masked datasets. 
Among the architectures tested, DeepSphere generally outperformed the pixel-based network across all tested resolutions and map types, suggesting its suitability for future high-resolution analyses. 
Both methods outperform the standard estimator for maps that contain relatively large $f_{\rm NL}$, as the estimator is optimized for $f_{\rm NL} = 0$. 
There is also only a limited performance drop on the recovered $f_{\rm NL}$ when we consider a wide range of $f_{\rm NL}$ versus $f_{\rm NL} = 0$.

While our models achieved (near-)optimal results in some settings, further refinements in training data volume and independence, network architectures, and hyperparameter settings will be critical for broader application. 
Specifically, larger, independent training sets that have been fine-tuned for the right resolution and additional parameter tuning may allow spherical CNNs to fully match or exceed the performance of traditional estimators at higher resolutions (nside 128 and above). 
Future work should also explore CNN-based extraction of non-Gaussian features beyond the bispectrum, assess models with realistic foregrounds and instrumental noise, and evaluate computational scaling for higher-order non-Gaussianity.

In this work, we also only considered one type of non-Gaussianity using only temperature maps of the CMB. 
It is known that polarization data can improve constraints on primordial non-Gaussianity in a non-trivial way when considering shapes that are not squeezed \citep{Kalaja2020mkq}. 
It would be interesting to explore how networks would perform on these other types of non-Gaussianity, and including polarization data. 
Likewise, although different shapes are distinguishable, we should check if trained networks are able to make this distinction by applying e.g. a network trained on local non-Gaussianity on a map containing equilateral non-Gaussianity. 

In summary, our findings suggest spherical CNNs could become a viable alternative in the search for primordial non-Gaussianity in the CMB, capable of efficiently handling high-dimensional map-level data. 
With continued development, CNNs have the potential to complement or even replace traditional methods, perhaps expanding our ability to probe inflationary models and fundamental physics from the early universe.

\section*{Acknowledgements}

The authors would like to thank Will Coulton, Adri Duivenvoorden, Eric Guzman, Joel Meyers, Joe Ryan, and Brandon Stevenson for useful discussions. We thank Brandon Stevenson for supplying us with additional code to run the KSW implementation from \citet{AdriKSW}, and Franz Elsner and Benjamin Wandelt for making their non-Gaussian CMB maps available to the public.





\bibliographystyle{mnras}
\bibliography{main} 




\appendix

\section{Full results}\label{app:full-results}

\FloatBarrier
\begin{table}
    \centering
    \begin{tabular}{|c||c|c|c|c|}
        \hline
        Model & \multicolumn{2}{c|}{DeepSphere} & \multicolumn{2}{c|}{Pixel-based}\\
        \hline
        \hline
        nside & Full range & Gaussian & Full range & Gaussian\\
        \hline
        \hline
        16 & 185$\pm$2 & 211$\pm$4 & 189$\pm$2 & 219$\pm$5\\
        \hline
        32 & 98$\pm$3 & 108$\pm$4 & 103$\pm$2 & 107$\pm$1\\
        \hline
        64 & 58$\pm$2 & 54$\pm$3 & 60$\pm$2 & 56$\pm$3\\
        \hline
        128 & 33$\pm$1 & 29$\pm$1 & 37$\pm$2 & 32$\pm$1\\
        \hline
    \end{tabular}
    \caption{Average results for both of the models that were trained on the unmodified dataset. The RMSE is given for the full range test set, containing $f_{\rm{NL}}$ uniformly distributed over the range $-1000$ to $1000$, and for a test set of Gaussian data.}
    \label{tab:normal-avg}
\end{table}

\begin{table}
    \centering
    \begin{tabular}{|c||c|c|c|c|}
        \hline
        Model & \multicolumn{2}{c|}{DeepSphere} & \multicolumn{2}{c|}{Pixel-based}\\
        \hline
        \hline
        nside & Full range & Gaussian & Full range & Gaussian\\
        \hline
        \hline
        16 & 208$\pm$1 & 242$\pm$6 & 221$\pm$2 & 250$\pm$6\\
        \hline
        32 & 107$\pm$1 & 119$\pm$5 & 111$\pm$1 & 122$\pm$3\\
        \hline
        64 & 61$\pm$1 & 57$\pm$4 & 62$\pm$2 & 64$\pm$4\\
        \hline
        128 & 34$\pm$1 & 31$\pm$2 & 41$\pm$4 & 37$\pm$2\\
        \hline
    \end{tabular}
    \caption{The same as in table \ref{tab:normal-avg} but now for the maps with noise.}
    \label{tab:noise-avg}
\end{table}

\begin{table}
    \centering
    \begin{tabular}{|c||c|c|c|c|}
        \hline
        Model & \multicolumn{2}{c|}{DeepSphere} & \multicolumn{2}{c|}{Pixel-based}\\
        \hline
        \hline
        nside & Full range & Gaussian & Full range & Gaussian\\
        \hline
        \hline
        16 & 224$\pm$1 & 250$\pm$4 & 229$\pm$4 & 257$\pm$5\\
        \hline
        32 & 112$\pm$2 & 130$\pm$2 & 125$\pm$3 & 133$\pm$4\\
        \hline
        64 & 68$\pm$2 & 64$\pm$2 & 70$\pm$3 & 69$\pm$2\\
        \hline
        128 & 38$\pm$1 & 34$\pm$1 & 42$\pm$2 & 37$\pm$1\\
        \hline
    \end{tabular}
    \caption{The same as in table \ref{tab:normal-avg} but now for the masked maps.}
    \label{tab:mask-avg}
\end{table}

In Tables \ref{tab:normal-avg}, \ref{tab:noise-avg} and \ref{tab:mask-avg} are the average results of all 5 runs trained on the normal maps without any modifications, trained on the maps with noise and trained on the maps with masking, respectively.
The results show a very similar pattern to the one observed in the best results from Tables \ref{tab:normal}, \ref{tab:noise} and \ref{tab:mask}.
The main difference is the average rmse over all runs is higher than the rmse of the best-performing model.

We also show the standard deviation on the average error.
While all runs are consistently able to learn to extract $f_{\rm{NL}}$, a difference in the test error of up to 13\% was observed between different runs of the same setting, both for DeepSphere and pixel-based models.
This is enough variation to warrant some discussion.
The standard deviation of the error obtained from training over the full range of $f_{\rm NL}$ is almost always smaller than the standard deviation on the Gaussian error.
This is to be expected, given that the CNNs are trained over the full range of $f_{\rm{NL}}$ and the loss is minimized over this entire range, which is the same type of performance measure as the rmse measured over the full range test set.
Current data suggest that any non-Gaussianity in the map is small. Hence it is especially the Gaussian error that is of interest to us.
The variation in these Gaussian errors suggests that our observed best models, likely are not the best possible models given the current parameter settings.
in other words,  even for the exact same CNNs, the best possible performance is likely closer to the optimal error than we have observed here.
This seems to especially apply to lower nside.
The standard deviation seems to shrink with nside and the overall magnitude of the average performance.

\begin{table}
    \centering
    \begin{tabular}{|c||c|c|c|c|c|c|}
        \hline
        Type & \multicolumn{2}{c|}{Normal} & \multicolumn{2}{c|}{Noise} & \multicolumn{2}{c|}{Masked}\\
        \hline
        \hline
        nside & Best & mean & Best & Mean & Best & Mean\\
        \hline
        \hline
        16 & -5 & -7$\pm$5 & 9 & -3$\pm$13 & -1 & -2$\pm$7\\
        \hline
        32 & 5 & 15$\pm$6 & 12 & 16$\pm$8 & 9 & 17$\pm$8\\
        \hline
        64 & 2 & -4$\pm$6 & -1 & -4$\pm$6 & -10 & -10$\pm$3\\
        \hline
        128 & 1 & 1$\pm$4 & 1 & -5$\pm$6 & 3 & -2$\pm$7\\
        \hline
    \end{tabular}
    \caption{The bias for the Deepsphere models over all types of data. Here we have given the mean bias over all 5 runs as well as the bias for the best-performing model, as measured by the Gaussian RMSE.}
    \label{tab:bias_deepsphere}
\end{table}

\begin{table}
    \centering
    \begin{tabular}{|c||c|c|c|c|c|c|}
        \hline
        Type & \multicolumn{2}{c|}{Normal} & \multicolumn{2}{c|}{Noise} & \multicolumn{2}{c|}{Masked}\\
        \hline
        \hline
        nside & Best & mean & Best & Mean & Best & Mean\\
        \hline
        \hline
        16 & 1 & 7$\pm$6 & -12 & -5$\pm$19 & 8 & 5$\pm$20\\
        \hline
        32 & 10 & 10$\pm$3 & 12 & 19$\pm$12 & 18 & 6$\pm$11\\
        \hline
        64 & 14 & 4$\pm$8 & -19 & -11$\pm$8 & 5 & 9$\pm$5\\
        \hline
        128 & -10 & -4$\pm$9 & -15 & -2$\pm$9 & 0 & -3$\pm$5\\
        \hline
    \end{tabular}
    \caption{The same applies as in table \ref{tab:bias_deepsphere}, but now for the pixel-based models.}
    \label{tab:bias_healpix}
\end{table}
Another possible issue we have only briefly discussed so far, is that the predictions of our models might be biased. 
To measure if the models indeed have a bias, we simply take the mean of the predicted values on the Gaussian test set.
The bias results for the Deepsphere models and the Healpix models have been given in tables \ref{tab:bias_deepsphere} and \ref{tab:bias_healpix}.
We show the bias of the best-performing model, as measured by the Gaussian error in Tables \ref{tab:normal}, \ref{tab:noise} and \ref{tab:mask}, as well as the mean bias over the 5 different training runs.
We find that most models have some small bias, albeit with a great variation in the bias between models. 
The main reason for this bias is likely the relatively low number of training points directly around $f_{\rm NL}=0$.
We expect that with a larger number of training points (around $f_{NL}=0$), the bias of the models gets increasingly closer to 0.
Furthermore, we expect that at higher resolutions, the tighter error margins will naturally push the bias to 0. 
This is also the trend we observe in our results, nside = 16 being the exception.

\FloatBarrier
\begin{table}
    \centering
    \begin{tabular}{|c||c|c|c|c|}
        \hline
        Type & \multicolumn{2}{c|}{Full range} & \multicolumn{2}{c|}{Gaussian}\\
        \hline
        \hline
        nside & Best & Mean & Best & Mean\\
        \hline
        \hline
        16 & 182 & 184$\pm$2 & 206 & 211$\pm$4\\
        \hline
        32 & 95 & 98$\pm$2 & 103 & 107$\pm$4\\
        \hline
        64 & 56 & 58$\pm$2 & 51 & 54$\pm$3\\
        \hline
        128 & 33 & 33$\pm$1 & 28 & 29$\pm$1\\
        \hline
    \end{tabular}
    \caption{Results of Deepsphere on the normal simulated CMB maps after debiasing. Aside from debiasing, the same conditions apply as in table \ref{tab:normal-avg}.}
    \label{tab:debias_norm}
\end{table}

\begin{table}
    \centering
    \begin{tabular}{|c||c|c|c|c|}
        \hline
        Type & \multicolumn{2}{c|}{Full range} & \multicolumn{2}{c|}{Gaussian}\\
        \hline
        \hline
        nside & Best & Mean & Best & Mean\\
        \hline
        \hline
        16 & 207 & 208$\pm$1 & 236 & 242$\pm$6\\
        \hline
        32 & 105 & 106$\pm$1 & 113 & 118$\pm$4\\
        \hline
        64 & 61 & 62$\pm$2 & 53 & 57$\pm$4\\
        \hline
        128 & 34 & 34$\pm$1 & 29 & 30$\pm$2\\
        \hline
    \end{tabular}
    \caption{Results of Deepsphere on the noisy CMB maps after debiasing. Aside from debiasing, the same conditions apply as in table \ref{tab:noise-avg}.}
    \label{tab:debias_noise}
\end{table}

\begin{table}
    \centering
    \begin{tabular}{|c||c|c|c|c|}
        \hline
        Type & \multicolumn{2}{c|}{Full range} & \multicolumn{2}{c|}{Gaussian}\\
        \hline
        \hline
        nside & Best & Mean & Best & Mean\\
        \hline
        \hline
        16 & 223 & 224$\pm$1 & 247 & 250$\pm$4\\
        \hline
        32 & 119 & 120$\pm$2 & 126 & 128$\pm$2\\
        \hline
        64 & 67 & 68$\pm$2 & 61 & 63$\pm$2\\
        \hline
        128 & 38 & 38$\pm$1 & 33 & 34$\pm$1\\
        \hline
    \end{tabular}
    \caption{Results of Deepsphere on the noisy CMB maps after debiasing. Aside from debiasing, the same conditions apply as in table \ref{tab:mask-avg}.}
    \label{tab:debias_mask}
\end{table}

Alternatively, we could remove the bias manually.
Inspection of the distribution of the predictions of the models on the Gaussian test set, reveals that the predictions follow a Gaussian curve.
This suggests the bias is not due to some imbalance in the predictions, but rather because of a shift in the mean of the predictions.
We can then remove the bias by simply subtracting the bias from all predictions the model makes.
Applying this procedure to the results of the Deepsphere models gives the results in tables \ref{tab:debias_norm}, \ref{tab:debias_noise} and \ref{tab:debias_mask}.
For both the Gaussian test set and, as well as in most cases the full-range test set, the predictions improved marginally.
Nevertheless, the effect of the bias seems to be very minimal on the RMSE of the models, which is likely another reason the models seem to struggle to remove the bias completely during training. 

\section{Test with large training set}\label{app:large-training-results}
\begin{table}
    \centering
    \begin{tabular}{|c||c|c|c|c|}
        \hline
        nside & 16 & 32 & 64 & 128\\
        \hline
        \hline
        Best RMSE & 204 & 98 & 47 & 26\\
        \hline
        Mean RMSE & 208$\pm$4 & 103$\pm$4 & 50$\pm$3 & 28$\pm$2\\
        \hline
        \hline
        Best RMSE debiased & 203 & 98 & 47 & 26\\
        \hline
        Mean RMSE debiased & 207$\pm$4 & 102$\pm$3 & 49$\pm$3 & 27$\pm$2\\
        \hline
        \hline
        Bispectrum & 189 & 93 & 47 & 22\\
        \hline
    \end{tabular}
    \caption{Results for the DeepSphere models that were trained on the dataset with more training examples on the Gaussian test set. Results are based on 5 runs per nside.}
    \label{tab:larger-training}
\end{table}
In this appendix, we describe an additional test that was conducted after the main results were obtained (see section \ref{sec:results} and appendix \ref{app:full-results} for these results).
The purpose of this additional analysis is to investigate the effect of an increased number of (independent) training maps on the results.
The main difference between this analysis and the ones conducted previously is the increased size of the training set as well as the number of independent seeds that were used to generate the training set.
Specifically, the training set consists of 40500 maps, generated from 900 of the independent $a^{T}_{lm}$ that were generated by \citet{elsner2009improved}.
This results in 45 maps per seed, all generated with a random $f_{\rm{NL}}$ between -1000 and 1000 and with a random rotation.
The remaining 100 $a^{T}_{lm}$ were used to generate the validation set, 10 maps per $a^T_{lm}$ again with random $f_{\rm{NL}}$ and rotations.
Note that we did not create a test set with random $f_{\rm{NL}}$.
Instead, we only use a Gaussian test set, the same test set we also used in the main analysis.

We limit our analysis to maps without noise and masking and only test the DeepSphere model, as we saw that for all maps the networks showed a similar trend in behaviour and the DeepSphere model performed better than the pixel-based networks in all tests we ran.
The same architecture and hyperparameter settings were used for the DeepSphere CNN in this test as in the main analysis described in Section \ref{sec:training}.

The results are displayed in Table \ref{tab:larger-training}.
We find that both the best error and the average error over all 5 runs are lower for every nside, as compared to the results from the main analysis in Tables \ref{tab:normal} and \ref{tab:normal-avg}, with further reduction after we remove any biases present in the models.
The use of more independent maps thus seems to improve the performance of the CNN.
Overall this analysis finds a similar trend in the behaviour as the main results, the error shrinking proportionally from nside 16 to 64 and increasing for nside 128. 
It is worth noting that already with the current training set, for nside $64$ the best model performs on par with the optimal bispectrum estimator.

An additional technique that could be used to improve the performance of models on higher nside, is to shrink the range of $f_{\rm{NL}}$ as we increase the nside.
The mean reason after all for using the large range of $f_{\rm{NL}}$ values is to prevent regression to the mean for low nside.
But for larger nside, we do not have to make this range as large to achieve the same effect.
As a test, we have generated an additional large training set for nside 128 in the same manner as discussed above, with the only difference being that we now take random $f_{\rm{NL}}$ values from a narrower range of $-200$ to $+200$.
The best-performing model has an RMSE of 25, while the mean RMSE over all models was 26$\pm$1.
Additionally, after removing the bias, the best model had an RMSE of 24 and the mean over all models was 25$\pm$1.
This again shows an improvement over all previous results, where we seem to get closer and closer to the optimal bispectrum error of 22.
This also suggests that a continued effort to improve and fine-tune the quality of the dataset (including the suggestions made here), can further improve performance.

\section{Flat experiment}\label{app:flat-results}
\begin{table}
    \centering
    \begin{tabular}{|c||c|c|c|c|}
        \hline
        Model & \multicolumn{2}{c|}{CNN} & Bispectrum\\
        \hline
        \hline
        Result Type & Full range & Gaussian & Gaussian\\
        \hline
        \hline
        Best RMSE & 85 & 85 & 90\\
        \hline
        Mean RMSE & 86$\pm$2 & 90$\pm$4 & 90 \\
        \hline
    \end{tabular}
    \caption{Results for the simplified flat map experiment. Here the maps have a resolution of 128x128 and the full range test set contains maps with $f_{\rm{NL}}$ values in the range -1000 to 1000. The optimal bispectrum error (at $f_{\rm{NL}}=0)$) is also given.}
    \label{tab:flat}
\end{table}
While true CMB simulations are inherently spherical in nature, we can create simplified flat maps that have similar properties.
We achieve this by perturbing a 2D flat grid in the same manner as the primordial curvature perturbations we describe in section \ref{sec:maps}.
To arrive at maps that are similar to the final spherical temperature CMB maps we have investigated this far, we follow the same steps, except using Fourier transforms instead of spherical transforms, thus performing the relevant operations in real and Fourier space where appropriate.
It is important to stress that the resulting maps are not a proxy of spherical CMB maps, but instead offer a simplified alternative problem that is relatively similar in nature.
Furthermore, these maps are thus not to be confused with flat projections of spherical maps. 
These maps were generated flat from the start.

The benefit of creating this simplified flat problem is that we can make use of standard flat 2D CNN algorithms that are very well-researched and heavily optimized.
This allows us to get a sense of the potential of future, more advanced spherical CNNs in analysing full sky CMB maps.
To test this potential, we apply a flat CNN that has a similar architecture to the spherical CNNs we have discussed in section \ref{sec:training}, albeit with some tweaks, such as double convolutions.
We apply this CNN to maps with a 128x128 resolution, which has a resolution that is similar to spherical nside 32 maps.
The results are given in table \ref{tab:flat}, where the best-performing model (as measured by the Gaussian error) has a bias of +8 and a mean bias close to zero. 
These results show that in the simplified flat scenario, even for low resolution we can consistently perform on par with the optimal estimator, both in the sense of the mean Gaussian error as well as the mean bias.
One thing to note is that the best-performing model appears to perform marginally better than the optimal estimator, which should be impossible.
This is likely a consequence of performing the analysis on the limited test set.
On an infinitely large test set, we expect the CNN to perform as well, but no better than the optimal estimator.


\bsp	
\label{lastpage}
\end{document}